# Exoplanet Exploration Program Analysis Group (ExoPAG) Study Analysis Group (SAG) 17 Final Report
# Resources Needed for Planetary Confirmation and Characterization

David R. Ciardi (Caltech/IPAC-NExScI), Joshua Pepper (Lehigh University),
Knicole Colon (Goddard Space Flight Center), Stephen R. Kane (University California Riverside)
With Input from the Astrophysical Community

## Table of Contents

# Executive Summary

The upcoming TESS mission will detect thousands of candidate transiting exoplanets. Those candidates require extensive follow-up observations to distinguish genuine planets from false positives, and to resolve the physical properties of the planets and their host stars. While the TESS mission is funded to conduct those observations for the smallest and most Earth-size candidate systems, the large number of additional candidates will have to be vetted and measured by the rest of the astronomical community. To realize fully the scientific potential of the TESS mission, we must ensure that there are adequate observing resources for the community to examine the TESS transit candidates and find the best candidates for detailed characterization. The primary purpose of this report is to describe the follow-up observational needs for planetary discoveries made by transit surveys - in particular TESS. However, many of the same types of observations are necessary for the other discovery techniques as well, particularly with regards to the characterization of the host stars and the planetary orbits. It is worth acknowledging that while a planet discovery may be a one-time event, the deeper understanding of a planetary system is an ongoing process, requiring observations with better precision over longer time spans.

# 1. Highlighted Points of Interest

1. Ground-based observations are a critical component to the success of the transit missions. Without the ground-based observations, the scientific goals of the missions can not be met. As such, the ground-programs are as significant to the missions as the spacecraft themselves.
2. Ground-based telescope resources are necessary to validate, confirm, and characterize exoplanets. Resources include a suite of telescopes spanning 1-m class through the 10-m class and capabilities must include a suite of instruments that enable wide-field imaging, high angular resolution imaging, spectrographs with resolutions of a few thousand or greater, and precision radial velocity spectrographs.
3. Financial resources to the community to support the necessary ground-based follow-up work of students, postdoctoral scholars, and early-career scientists are needed.

# 2. Introduction

The K2 and TESS missions have and will produce a new population of planets to study. These discoveries begin as candidates, and are then confirmed as planets, and after that require additional analysis to derive reliable system parameters, including the masses and radii of the planet, the planet orbital properties, the characteristics of the host star, and the presence of other stars or planets in the system.

Transit surveys are known to be subject to false positives (Brown 2013). The determination of false positives is closely tied to the confirmation of planets. Confirmation usually requires either a dynamical measurement of the candidate object's mass, typically through radial velocity orbital measurements, or a clear upper limit on the companion's mass, as through Doppler tomographic confirmation (Collier Cameron 2009). An alternative to confirmation is the process of "validation", taken to mean a thorough statistical analysis of the candidate to demonstrate that its planetary nature is statistically likelier than any of the known false positive configurations beyond some threshold (Morton 2012). While validation is a useful tool for ensemble statistical analysis of a planetary population, we concentrate on the process of confirmation.

While there are a few ways to dynamically confirm a transit candidate, the expectation for TESS is that the vast majority of the candidates will be confirmed via orbital RV spectroscopy. With unlimited capacity for high-precision RV observations, all candidates could be investigated this way, but in practice precision

RV observations are a scarce resource, and so an iterative approach using cheaper techniques is usually taken.  In that process, once a candidate is identified, it is observed with successively more expensive or resource-intensive methods, each of which is designed to exclude some type of known false positive.  As the candidate passes each test, it is handed off to the next observational step, until most common types of false positives are excluded, and precision RV can be employed only on the remaining candidates for orbital determination and confirmation.

After the planetary nature of a candidate is confirmed, the work is not done.  Knowing that there exists an exoplanet orbiting a particular star is generally not that useful unless we can reliably describe the system parameters.  Those include the temperature, mass, radius, metallicity, age, rotational velocity, and space motion of the host star, along with membership in any cluster, association, or galactic population.  For the planet, it includes the mass and radius, along with orbital properties (semimajor axis, eccentricity, inclination, spin-orbit alignment, longitude of periastron) and also potentially atmospheric properties such as albedo, the day-side thermal emission and the amount of atmospheric redistribution.  Many of these properties are determined in the process of detecting or confirming the planet, but not all of them.

One of the most crucial needs for measuring the planet properties is the identification of all nearby luminous sources that contribute flux to the light curve of a transit signal, whether in the original survey observations or in follow-up observations.  If such sources are not accounted for, the calculated planet size will be incorrectly determined, and will generally be underestimated.  If a star with a transiting planet is seen in a survey light curve, but another fainter star is located in the target star aperture and is not accounted for, the blended star (whether physically associated with the target or not) will dilute the depth of the transit, leading to an underestimation of the planet radius.  Likewise, even if the target star is not blended with another object, it is necessary to have an accurate measurement of the host star radius to accurately determine the planet radius.

## 3. Expected Number of TESS Planetary Candidates

TESS, launched in April 2018 and expected to start operations summer 2018, will observe around 200,000 stars at a 2-minute cadence using selected postage stamps on the pre-selected targets, and will observe at least 2,000,000 stars at a 30 minute cadence in the full-frame images with enough photometric precision to detect planetary transit signals.  From the postage stamps alone, simulations have shown that TESS is expected to find ~5000 transit signals, about 35% of which are expected to be real planets for a total yield of ~2000 planets.

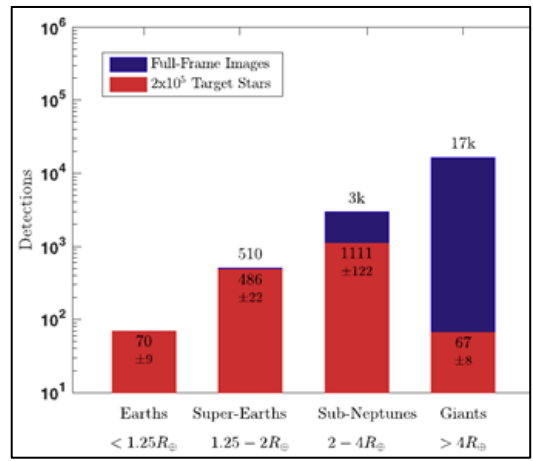

From the full frame images, the number of transit detections is expected to be an order of magnitude larger with more than ~50,000 candidates and assuming a similar fraction of planetary detections resulting in ~20,000 planets. In Figure 1, the numbers of planets expected to be detected by TESS from the postage stamps and the full frame images are shown (Sullivan et al. 2015).

*Figure 1: Expected planet yield as estimated by Sullivan et al. (2015).  The simulations predicted ~2000 planets from the 2-minute postage stamps and ~20,000 planets from the 30-minute full frame images.*

Additional studies by other groups since this original simulation have found similar numbers (although the studies differ in detail) for the expected yield of TESS (e.g., Bouma et al. 2017; Barclay, Pepper, & Quintana 2018).

# 4. Planet Confirmation & Exclusion of Astrophysical False Positives

## 4.1. Planet Confirmation

Transit surveys are subject to a range of false positives. This is especially true of ground-based transit surveys, due to the types of planetary and orbital configurations they probe. Because such surveys are much more sensitive to large planets on short orbital periods, they are primarily able to detect Hot Jupiters. It is now known that Hot Jupiters are intrinsically rare (Wright et al, 2012), and many different configurations of eclipsing binaries (EBs) can mimic the photometric signature of a planet (Brown 2003; Torres, 2004; O'Donovan, 2006). Transit surveys from space have greater duty cycles and photometric precisions, allowing them to detect longer period and smaller planets, for which fewer standard false positive configurations occur, but which may still be present.

Each type of false positive can require a different type of observational check, spanning observational techniques including both single-epoch and time-series spectroscopy, and single-epoch and time-series photometry of varying special resolutions. Here we describe the most common types of false positives with the methods and observational resources for identifying them.

## 4.2. False Positive Types

**Configuration #1: Eclipsing Binary**: An eclipsing binary with a large primary star and a small secondary star, such as an M dwarf orbiting an F dwarf or a grazing eclipsing binary.

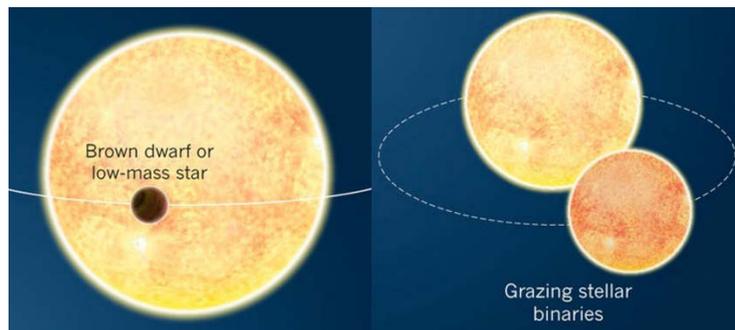

Signal: An M dwarf is roughly the size of Jupiter, so the primary eclipse of the M dwarf in front of an F or G dwarf has a depth comparable to that of a giant planet transit in front of a dwarf star. A primary eclipse by a late-type dwarf star in front of a giant star has a depth comparable to that of an earth/super-earth in front of a dwarf star. Also, a grazing eclipsing binary of any type can create an eclipse of any apparent depth.

*Figure 2: Eclipsing binaries blended with foreground and/or background stars can appear as planetary transiting systems to transit surveys where the primary target actually has a brown dwarf or stellar companion..*

Resolution(s):

- Grazing eclipsing binaries usually have eclipse shapes quite different from full eclipses/transits, and can generally be identified in the survey light curves.
- Detection of a secondary eclipse in the discovery photometry can demonstrate that the transiting object is stellar.
- Spectroscopic observations of the star can determine the luminosity class of the primary and therefore determine if the transiting object has a radius more akin to a planet or star.
- If the secondary star is bright enough, a single RV observation can detect the blended light of the secondary (SB2).
- Multi-epoch radial velocity observations are the most critical observations as these can directly determine if that the mass of the secondary is stellar. At minimum, two RV observations near opposite quadrature are required (SB1).
- An additional technique of multi-color time series photometry of the transits can be used to search

for a chromaticity dependence in the transit depths indicative of the self-luminosity of a stellar (non-planetary) companion; however, for very small transit depths, this is not always feasible from the ground. Chromaticity checks benefit from the availability of observing in a broad range of wavelengths from the optical through the infrared.

Complication(s)/Limitation(s): Detection of the secondary eclipse requires sufficient photometric precision, as does multiband time-series photometry for chromaticity measurements. RV observations require that the primary star be bright enough for RV measurements and that the radial velocity signature be of sufficient amplitude to be detectable.

**Configuration #2: Eclipsing Binary Blended with Brighter Star**: An EB blended with a brighter star, either line of sight or in a hierarchical triple.

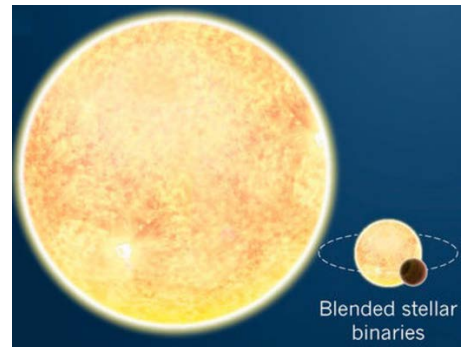

*Figure 3: Similar to Configuration #1 but the blended eclipsing binary is not the primary target but a background/foreground binary that was not previously known to be there.*

Signal: The brighter star can dilute the measured depth of the primary eclipse to that of a planet, and the secondary eclipse to below the level of photometric detection.

Resolution: A key step here is to assess whether another star is blended with the candidate star in the photometric aperture of the discovery light curve. If so, it must be determined whether the "neighbor" star is of sufficient brightness for a blended eclipse of that star to mimic the observed transit in the discovery light curve. If the neighbor star is sufficiently separated from the (supposed) host star in angular space, several techniques can be employed:

- The centroid of light during transit can shift from the neighbor to the primary target if the neighbor is an EB, which can be detected in the survey photometry.
- Additional time-series photometry with higher angular resolution can be used to directly observe the occurrence of the eclipse event on the neighbor star.
- Radial velocity observations can be made of the two stars depending on the angular separation of the two stars.
- Multi-color single-epoch high resolution imaging is the most critical of the observations to ascertain whether or not the primary target is blended with another star, and what the properties of that blending star are, including whether it is likely bound or unbound, and what effect that companion star has on the derived planetary parameters.
- If the neighboring star and the target star are blended too closely for observations to angularly resolve the stars, a deep spectrum may be able to identify the presence of a "companion" star and be used to determine either the spectral type of the neighbor or its approximate apparent magnitude.

Complication(s)/Limitation(s): See Category #1 above for the photometric and RV methods. For the RV observations, if an effort is made to identify the spectrum of the fainter object, sufficient sensitivity is required. High angular-resolution imaging must have sufficient angular resolution for identification of closely blended objects, and sufficient sensitivity to detect blended objects for which a total eclipse can produce a signal comparable to the discovery photometric signal when blended with the target star.

**Configuration #3**: **Planet-Host Star Blended with Star**: A star with a giant planet blended with a brighter star, either line of sight or in a hierarchical triple.

Signal: The companion star can dilute the measured depth of the transit such that the derived planetary radius is smaller than the true planetary radius. This scenario is similar to Blended Eclipsing Binary (Configuration #2) except that the eclipsing body is planetary and not stellar.

Resolution: This configuration can be identified by the same tools as for Configuration 2, although spectroscopic observations likely will not be able to identify the star with a planet in the manner of an SB1 or SB2.

Complication(s)/Limitation(s): See Category #2 above.

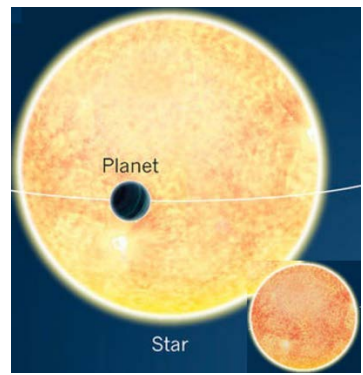

*Figure 4: A true planet hosting star but a blended star has diluted the transit to make the transit appear shallower than in reality.*

**Configuration #4: Spurious Single Star**: A single, isolated star with a spurious transit signal.

Signal: Between intrinsic astrophysical variability and systematic noise, spurious identifications of transit candidates are possible, and indeed, expected, especially when searching for small planet transits.

Resolution: For ground-based surveys, follow-up observations with larger telescopes that can achieve high photometric precision can confirm that there is an astrophysical transit signal at the expected ephemeris. For space-based missions like TESS and K2, it is generally not possible to obtain higher photometric precision from the ground than in the discovery light curve. If additional space-based photometric resources are available (HST, Spitzer, CHEOPS, etc) those can be employed, but are expensive. Therefore, dynamical RV orbital measurements are needed to measure the mass of the candidate planet, or at very least place upper limits on the mass.

Separately, single-epoch spectra or a handful of spectra can identify the signature of chromospheric activity that can suggest high rotation, spot modulation, or other signs of intrinsic astrophysical variability that could be the cause of the candidate transit signal. Additional single-epoch or time-series photometry in different bandpasses than the discovery light curve can also detect such signals.

Complication(s)/Limitation(s): This category is the most difficult one, because it is possible to expend substantial observing time and effort to confirm a signal that is actually spurious. The most important issue here is to identify a series of criteria before follow-up is undertaken to decide when to call a detection "likely", "unlikely", or "ambiguous", and cease additional observations. It is also strongly dependent on the SNR threshold selected for identifying candidates in the survey photometry.

### 4.3. Recommended Approach and Considerations

To create a process for dealing with FPs, it is best to have a "recipe" that a follow-up management office can follow. That recipe would involve a continuous evaluation of each transit candidate, starting from the point that it is designated as a candidate by whatever survey office has that responsibility. The recipe would take into account the possible types of known false positives that could be present in the candidate, and would then assign a likelihood to each scenario. The likelihood number does not have to be an actual statistical likelihood of the scenario, but is rather used to prioritize which kind of follow-up observations are worthwhile. That prioritization would also incorporate the cost of the observations in time or money or both, along with the intrinsic value or importance of the candidate, if real.

The exact procedure for the prioritization of the candidates and the assignment of follow-up resources is not within the scope of this report. Rather, the important issue is that there be a process, and that it be centrally managed, yet flexible. It does not require full control over the follow-up resources, but instead it should provide guidance for the mix of participants that will be participating in the follow-up observations. It is also important that the follow-up management entity synthesize the results of the observations as much as is practical to take the following steps:

- Verify whether the target is indeed a false positive, of what kind, and based on what observations.
- Confirm that a given false positive scenario has been reasonably excluded.
- Re-prioritize each type of follow-up observation after new data is acquired.
- Revise the observational or inferred physical properties of the candidate star and planet(s).
- Declare a candidate "confirmed" or "verified".

A sample recipe for false positive elimination could look like this:

1. Single epoch spectroscopy of all candidates. Identify and eliminate SB2s. Verify that the host star parameters are correctly listed in the input catalogs, and that the derived transit properties are consistent with a planetary companion. Adjust the stellar properties as listed in the ExoFOP[1] site as needed.
2. Multiple epoch spectroscopy at medium RV precision of all candidates that pass the first round. Observations at predicted orbital quadrature can identify and eliminate SB1s.
3. High angular resolution single-epoch photometry of all candidates that pass rounds 1 and 2. Candidates with no significant astrophysical sources blended in the discovery photometry aperture can then be passed to high-precision RV observations for orbital determination.
4. Candidates that pass rounds 1 and 2 but are shown to be significantly blended with nearby objects will require additional vetting. Seeing-limited time-series photometry can be conducted to identify whether the transit signal originates from the target star or a neighbor.

Candidates that pass all rounds above can be sent for precision RV observations to determine orbital parameters.

### 4.4. Planet Confirmation

As noted, transit candidates may be confirmed as planets through a variety of methods, all of which determine the mass of the transiting body, or place an upper limit on the mass. Although an orbital determination of the radial-velocity signature of the transiting object technically provides only a minimum limit on the mass, if the object is transiting the host star, the inclination is constrained enough such that the factor of $\sin(i)$ is typically very close to unity.

The ability to perform an RV confirmation of a given candidate depends on both astrophysical factors and instrumental capability. The astrophysical factors include:

- The apparent brightness of the host star
- The temperature of the host star and availability of spectral features
- The luminosity class of the host star
- The metallicity of the host star
- The projected rotation rate of the star
- The amount of chromospheric activity

The instrumental factors include:

---

[1] https://exofop.ipac.caltech.edu

- The effective aperture of the telescope
- The resolution of the spectrograph
- The wavelength range
- The stability of the spectrograph
- The wavelength calibration method
- The throughput efficiency

There are additional methods, using the light curve photometry, that can be used for planetary confirmation, although they tend to be suitable for only a limited set of star-planet properties and configurations. They include:

- Ellipsoidal variations
- Reflected or emitted light phase curves
- Doppler boosting

Such methods are generally only employed in small numbers of special cases, and need not be considered for the purposes of a comprehensive follow-up approach.

### 4.5. Ephemeris Recovery

While the discussion above deals with the need to confirm the planetary nature of candidates and derive their parameters, there is one crucial piece of information that is different from the rest, which is the transit ephemeris. While quantities like stellar radius and temperature or planet radius do not change with time, the ephemeris of a planet grows less certain as time passes since the most recent transit observations. After a significant number of later orbits have passed with no re-observation, the ability to predict future transits is completely lost. Even for systems with 2 or 3 orbits observed in the discovery light curve, the ephemeris may degrade over the course of 6 months after the discovery observations to worse than several hours.

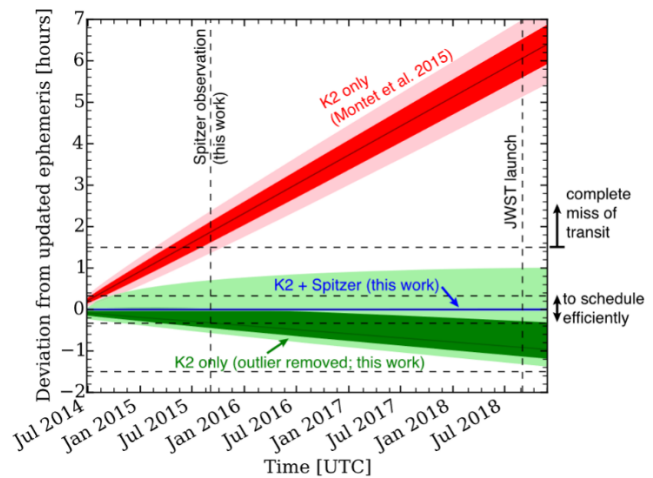

Figure 5: Ephemeris accuracy of K2-18b with K2 only (red) and with K2+Spitzer follow-up (green) projected out to the original October 2018 launch date of JWST. With the delay of JWST launch until 2021, the accuracy and precision of TESS transits is even more critical (Benneke et al. 2017).

For later observations of the transits with facilities like HST or JWST, it will generally be necessary to know the upcoming transit time to better than 1 hour precision. Thus, later observations that can establish more precise ephemerides will often be required. This phenomenon has already been seen with planets discovered by K2, as noted by Benneke et al. (2017) and Stefansson et al. (2018). Planets with only single transits seen in the discovery light curves will have even more poorly determined ephemerides.

 Later observations to fix transit ephemerides can involve both time series photometry (Figure 5; e.g., Benneke et al. 2017 & Stefansson et al. 2018), or RV observations. The contours of this problem have not been examined in detail yet, but multiple groups are investigating how many of the TESS planets will require later photometric followup, the type of planets that will most likely have completely lost ephemerides, the role of RV observations, and amount of resources needed for the expected observations. Any accounting of resources needed for confirming TESS planets will need to include ultimately the resources needed for ephemeris rescue, especially if the potential of JWST observations of TESS planets is to be realized.

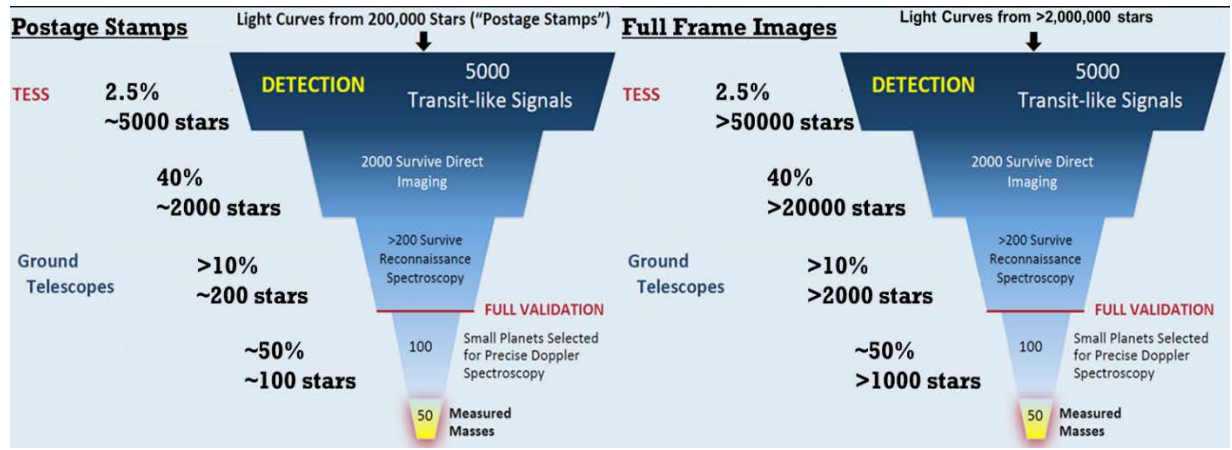

*Figure 6: Hierarchy of the required follow-up showing the estimated number of targets that will need observing. The inverse pyramid shows the progression through the various types of observing described below including seeing-limited imaging, spectroscopy, high-resolution imaging, and precision radial velocities (figure adapted from TESS).*

## 5. Resources Needed

In this section, we review the amount of resources estimated to be necessary for the validation, confirmation, and characterization of the planetary candidates to be discovered by TESS. There is a hierarchy of follow-up that is necessary starting with seeing-limited time series observations, spectroscopy and high resolution imaging – all of which is necessary to validate the planetary candidates. After validation, confirmation with precision radial velocities is necessary to obtain the planetary masses. Each step yields a progressively smaller but more reliable planetary candidate sample that require increasingly rarer and more precious telescope resources (see Figure 6).

Without the resources for dedicated follow-up and characterization, the bulk of the science yield of TESS will remain unfulfilled. Resources required for the necessary identification of false positives, the validation and confirmation of exoplanetary candidates and the characterization of the planets include seeing-limited time-series photometry, stellar spectroscopy, high-resolution imaging, and precision radial velocity. The time requirements for each of these techniques are estimated in the sub-sections below for the expected number of candidates from the TESS postage stamps and the TESS Full Frame Images – with a summary given in Table 1 at the end of this section.

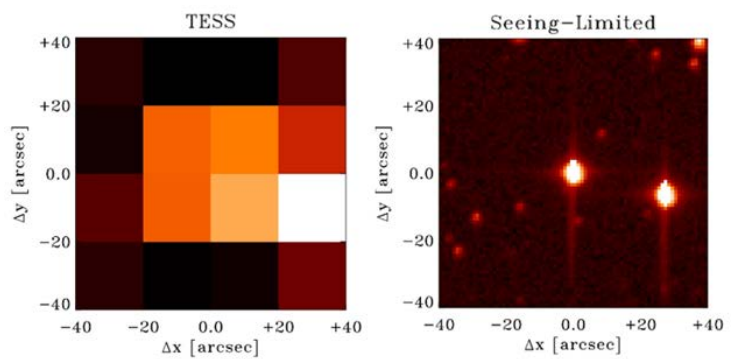

### 5.1. Seeing-limited Time Series Photometry

*Figure 7: Simulation of the TESS view of Kepler-1002, which has a bright star 20" away. For TESS, these two stars are blended and seeing-limited time series photometry would be necessary to show that the transit event is not an eclipsing binary signature on the star to the east.*

With the relatively large pixels of TESS (21"), the primary contaminate of the planetary target candidates will be blended eclipsing binary (BEBs) stars (Figure 7). Seeing-limited time-series photometry will be the first line of defense in determining if the observed transit is caused by a nearby blended eclipsing binary (e.g., Collins et al. 2018). The observations need not detect the planetary transit itself, but rather need to determine if the TESS-detected signal is in fact due to a nearby EB. This kind of observation can be performed with relatively small telescopes (< 1m), and ~1% precision photometry is needed. The observations also do not need to fully sample the eclipse/transit time, but rather a series of observations prior to the transit, during the transit, after the transit may be sufficient.

With typical observation times of 15-30 minutes per star, we estimate that ~5,000 – 10,000 hours of observation are needed for the postage stamp candidates and >10,000-20,000 hours of observation are needed for the full frame image candidates. If uninterrupted observations throughout the transit events are required, then the time estimate becomes substantially larger.

## 5.2. Stellar Spectroscopy

Determination of the stellar properties, in particular the stellar radius, is critical to the determination of the planetary radius. Because the measured transit depth from the light curves is the ratio of the planetary radius to the stellar radius (squared), accurate and precise determinations of the stellar radii are necessary. Fulton et al. (2017) showed how crucial this was with the discovery in the Kepler planetary radius distribution of a radius gap near a radius of 1.8 $R_{Earth}$ (see Figure 8).

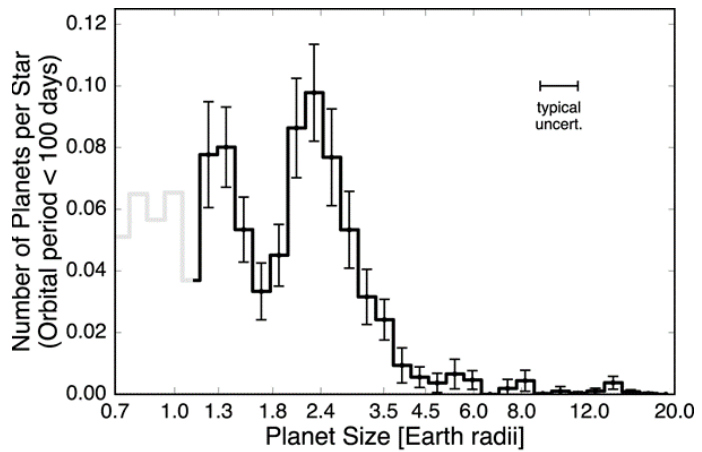

Figure 8: Gap in the radius distribution of planets that only became visible with precise and accurate stellar radii (Fulton et al. 2017)

To obtain the stellar radius, one can use stellar spectroscopy to determine $T_{eff}$, and combine that with the distance to the star from Gaia (Stassun et al. 2018). However, spectroscopy, along with stellar isochrone models, is still needed to determine the metallicities of the stars, which are necessary for the stellar mass determinations. Spectroscopy can also provide the rotational velocities of the stars, which are necessary for assessment of the suitability for radial velocity observations.

Stellar spectroscopy observations typically require spectrographs of resolutions spanning 5000 - 100,000 on 1 - 4m class telescopes. Both optical and infrared spectrographs are necessary, as many of the transiting host targets are late type K and M stars which are significantly brighter in the infrared than they are in the optical (e.g., Ciardi et al. 2018). For stars that survive the seeing-limited time-series vetting, spectroscopic observations will typically take 5 - 10 minutes per observation per star. We estimate that ~1,000 hours of telescope time is needed to observe the postage stamp candidates, and more than >10,000 hours to observe the full frame image candidates.

## 5.3. High Angular Resolution Imaging

Full validation and vetting of the transiting candidates requires imaging with high resolution to identify those stars that are multiple star systems. High resolution imaging has become the standard for determining the photometric blending caused by bound companions, and hence, determining the true planetary radii (e.g., Ciardi et al. 2015; Hirsch et al. 2017, Furlan et al. 2017 and references therein).

By sampling across the stellar companion period (separation) distribution (e.g., Raghavan et al. 2010, Kraus et al. 2016), speckle and adaptive optics observations in both the optical and near-infrared are necessary for the full validation and characterization of the planetary companions (Figure 9).

On 3-10m class telescopes, optical and near-infrared speckle and adaptive optics observations have resolutions of 0.01" – 0.1" and typically require 5 – 10 minutes per observation. Similar to the stellar spectroscopy observations, we estimate that ~1,000 hours of telescope time are needed to observe the postage stamp candidates, and more than 10,000 hours would be needed to observe the full frame image candidates.

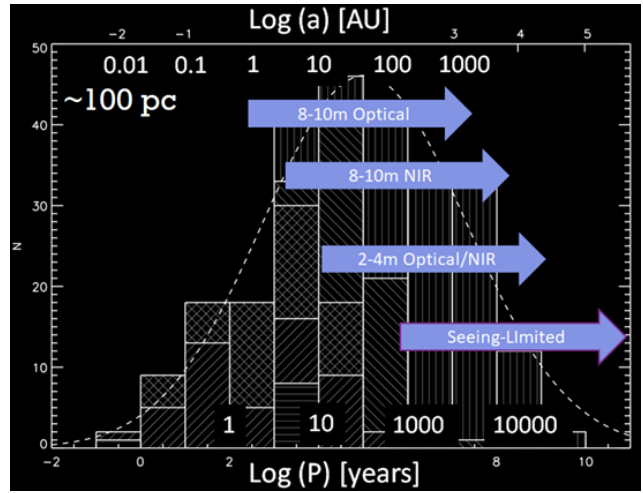

Figure 9: Sampling of the stellar companion distribution by high resolution imaging. The period distribution is from Raghavan et al. (2010) and was converted to physical separations assuming the stars are 100 parsecs away. The closer the stars get; the more powerful the high resolution imaging becomes.

### 5.4. Precision Radial Velocity Spectroscopy

Planetary confirmation, in comparison to validation only, requires the determination of which star the transiting body orbits and the determination of the mass of the orbiting body. Although other techniques have been used for confirmation (e.g., transit timing variations), precision radial velocities are the most utilized resource to ascertain that the planet indeed orbits a specific star and to determine the mass of the orbiting body.

Via Doppler tomography and/or the Rossiter–McLaughlin effect (Figure 10), precision radial velocity observations during the primary transit can be used to confirm the presence of an orbital body around the target star and ascertain the relative alignment of the planet orbit with the stellar rotation axis. Additionally, of course, precision radial velocity observations across the orbital periods of the planets are utilized to obtain the orbital parameters and masses of the planets (Figure 11).

Typical precision radial velocity observations require precisions of ~1-10 m/s, and each observation takes 5-30 minutes on 1-10m class telescopes. The number of observations required depends strongly upon the complexity of the system. For a single planet orbiting a quiet star, the number of radial velocity observations is typically a few spread across the planetary orbital period, but for multi-planet systems or for systems with active stars, many hundreds of observations are often needed. We estimate that approximately 2,000 hours of telescope time are needed to observe the postage stamp candidates, and more than 10,000 hours to observe the full frame image candidates. Additionally, special attention needs to be paid to how the observing is scheduled and how resources from different observatories are combined in order to best sample the expected orbital periods and stellar activity (Figure 11).

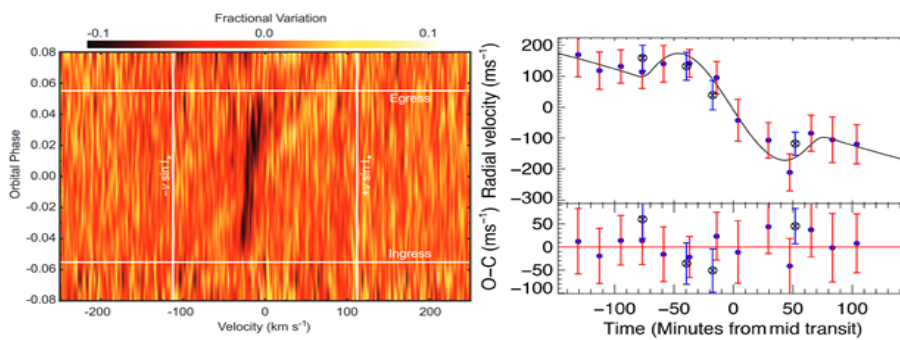

Figure 10: Doppler tomography image of KELT-9b (Gaudi et al. 2017) and Rossiter-McLaughlin effect measurements of WASP-103b (Addison et al. 2016). Both techniques require precision radial velocity observations and are used to confirm that the transiting body orbits the observed star and to determine the relative alignment of the planetary orbit with the stellar rotation axis.

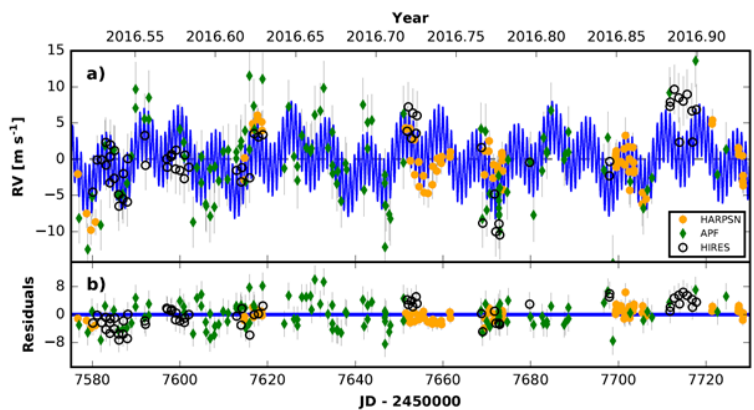

*Figure 11: Radial velocity observations of the 3-planet system HD 3167. Precision radial velocity observations were required from three different observatory longitudinal locations over the course of half a year in order to obtain the masses and orbits of the three planets (Christiansen et al. 2017).*

### 5.5. Archival Data Resources

New observations are needed for the validation, confirmation, and characterization of the planetary candidates, but the use of archival data is also critical. Examples of the use of archival data include: all sky imaging used to assess the presence of background sources revealed by the proper motion of the nearby stars (Figure 12); Gaia data enabling searches for nearby bound companions (e.g., Ziegler et al. 2018); Gaia astrometry used to find very-close in binary stars and even potentially astrometrically detect some of the larger planets.

Another important resource for the community is the ability to share data and information about observations taken and results obtained. This has been highly successful for Kepler and K2 where the ExoFOP website has been used by both the project and the community to record the observations that have been taken and the data and commentary resulting from those observations.

Such a sharing environment is necessary to prevent the acquisition of duplicate observations on the same targets and wasting precious telescope resources. As indicated above, the number of telescope hours necessary for each of the vetting steps is quite large (hundreds to thousands of hours), and the efficient and effective use of the telescopes requires the community to know what other people have already done (and shared). The TESS project is utilizing the ExoFOP website to communicate amongst itself and to the community what follow-up observations have been done. The community, as a whole, is continuing to utilize

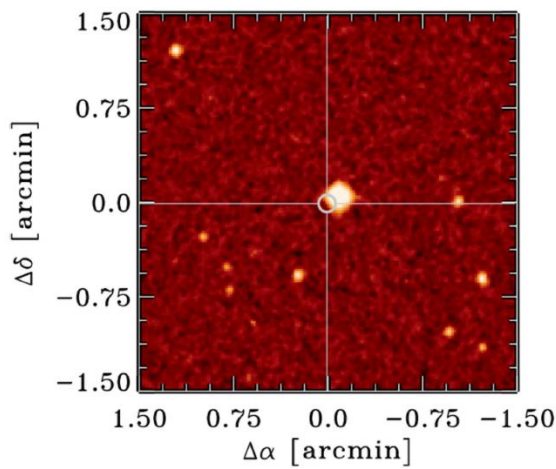

*Figure 12: Palomar Sky Survey image from 1955 showing the current position of K2-136 is not centered in front of a background star. Proper motion imaging was used as part of the validation process for the transiting planet (Ciardi et al. 2018)*

the ExoFOP-Kepler and ExoFOP-K2 portions of the site and we expect that to continue for the ExoFOP-TESS.

| Observing Technique | Typical Telescope Size | Number of Postage Stamp Candidates | Hours Needed for Postage Stamp Candidates | Number of Full Frame Image Candidates | Hours Needed for Full Frame Image Candidates |
|---|---|---|---|---|---|
| Seeing-Limited Time Series Photometry | < 1.0 m | ~5000 | ~5000 – 10,000 | >50,000 | >50,000 – 100,000 |
| Stellar Spectroscopy | 1 – 4 m | ~2000 | ~1000 | >20,000 | >10,000 |
| High Resolution Imaging | 4 – 10 m | ~2000 | ~1000 | >20,000 | >10,000 |
| Precision Radial Velocity | 1 – 10 m | ~200 | ~2000 | ~2,000 | >10,000 |

*Table 1: Summary of the Estimated Minimum Number of Telescope Hours that are needed to observe the TESS Planetary Candidates*

## 6. Comparison to the Kepler and K2 Follow-Up Observation Programs

The Kepler Mission observed about 200,000 stars and produced approximately 10000 Kepler Objects of Interest – of which, about 4500 were thought to be real planetary candidates that required follow-up for validation and confirmation. The Kepler Follow-Up Observation Program (KFOP) consisted primarily of spectroscopic observations for determination of stellar parameters (Furlan et al. 2018) and high resolution imaging for identification of nearby stellar companions (Furlan et al. 2017). Nearly all of the candidate KOIs were observed spectroscopically and with high resolution imaging. Additionally, radial velocity observations were made of a select set of KOIs that were amenable to mass determinations either because the semi-amplitude signal was expected to be large (> 1 m/s) and/or the stars were bright (Kepmag < 14$^{th}$ mag).

There were 6 main groups that were funded by the project during the primary mission: 3 groups were tasked with obtaining spectroscopic parameters and radial velocity observations and 3 groups were tasked with obtaining optical and near-infrared high resolution imaging. In addition to the funded program, there were many community members who contributed unfunded observations to the follow-up of the KOIs.

It is difficult to provide an accurate value for the total number telescope-nights that were used in the follow-up of Kepler targets, but we estimate that the funded Kepler FOP spent approximately 1000 telescope-nights from 2009 – 2015, averaging about 150 nights per year. When incorporating the time contributed by the general community over the same time period, we estimate that between 2000 and 3000 telescope-nights have been used by the community to follow-up Kepler targets.

For K2, because there was no centralized follow-up program but rather individual collaborations that worked together or often competed against each other, it is more difficult to assess the number of telescope-nights that have been used to follow-up K2 candidates. Based upon a literature search of the published K2 confirmed/validated planets and the published K2 catalog papers, we conservatively estimate that a few hundred telescope-nights per year have gone into the high resolution imaging, spectroscopy, and radial velocity follow-up – leading to 1000 – 2000 telescope-nights over the past 5 years since the start of K2 – some of which has been duplicated between competing groups.

Given that TESS will observe a similar number of targets in the postage stamp mode as Kepler and K2, it is reasonable to assume that TESS follow-up will require at least 1000 – 2000 telescope-nights just for the short cadence targets. When the full frame images are considered and the total number of candidates that may emerge from the full frame images, the amount of telescope time necessary to validate or confirm all of the candidates will exceed more than 10000 telescope-nights.

## 7. Financial Challenges for the Community

There are a variety of challenges which face the community in its effort to confirm and characterize planetary candidates found by TESS and find the best candidates suitable for detailed characterization with ground-based facilities and/or space-based facilities such as HST, JWST, and, eventually, WFIRST, or LUVOIR or HabEx.

One of these challenges that is often overlooked is financial. The primary group of scientists responsible for the observing, the data analysis, and the scientific discoveries are (and will be) students and postdoctoral scholars. Telescope observing time generally does not come with funding to support travel to the telescope or to support the analysis of the data. The exception to this is the NASA funded time on Keck and NN-Explore, but these funds are typically only sufficient to support travel and are not enough to support the analysis of the data once they are collected.

The funding available within the NASA Exoplanet Research Program is highly competitive, oversubscribed, and covers all of exoplanet research. As a result, the NASA Explanet Research Program will likely only be able to support 1 – 3 programs per year – assuming those programs are found to be more compelling than other proposed exoplanet science that does not involve follow-up observations. Further, the TESS GI program only allows for <30% of the program to be ground-based follow-up. As a result, the financial support necessary for the students and post-doctoral scholars to obtain and analyze the data necessary to enable the full scientific reaping of the missions dedicated to find exoplanets suitable for detailed characterization is lacking.

A significant effort in the determination of the exoplanet characterization is the preparatory and follow-up observation work that must accompany the candidate discoveries before the planets are ever confirmed or characterized in detail. Multi-million dollar annual support (competitively selected) is needed for the community to do all of this work and make the planetary discovery missions the success they are expected to be.

We estimate that an additional dedicated ~$1-2M dollars per year over the next 2 – 4 years could fund approximately 5 – 10 programs per year depending on the mix of postdoctoral scholars and students within each program. Such a dedicated effort would augment effectively the efforts of the funded TESS explorer team and enable sufficient (if not complete) follow-up to validate and confirm TESS candidates and enable informed target selection for more detailed characterization with JWST.

# References

Addison, B. C. et al. 2016, "Spin-orbit alignments for Three Transiting Hot Jupiters: WASP-103b, WASP-87b, & WASP-66b," ApJ, 823, 29

Barclay, T., Pepper, J., & Quintana, E.V. 2018, "A Revised Exoplanet Yield from the Transiting Exoplanet Survey Satellite (TESS)", arXiv:1804.05050

Benneke, B., Werner, M., Petigura, E., et al. 2017, "Spitzer Observations Confirm and Rescue the Habitable-zone Super-Earth K2-18b for Future Characterization" ApJ, 834, 187

Bouma, L. G., Winn, J. N., Kosiarek, J., & McCullough, P. R. 2017, "Planet Detection Simulations for Several Possible TESS Extended Missions" arXiv:1705.08891

Brown, T. M. 2003, "Expected Detection and False Alarm Rates for Transiting Jovian Planets", ApJL, 593, L125

Christiansen, J. L. et al. 2017, "Three's Company: An Additional Non-transiting Super-Earth in the Bright HD 3167 System, and Masses for All Three Planets," AJ, 154, 122

Ciardi, D. R. et al. 2015, "Understanding the Effects of Stellar Multiplicity on the Derived Planet Radii from Transit Surveys: Implications for Kepler, K2, and TESS," ApJ, 805, 16

Ciardi, D. R. et al. 2018,"K2-136: A Binary System in the Hyades Cluster Hosting a Neptune-sized Planet," AJ, 155, 10

Collins, K. A., Collins, K. I., Pepper, J., et al. 2018 "The KELT Follow-Up Network and Transit False Positive Catalog: Pre-vetted False Positives for TESS", arXiv:1803.01869

Fulton, B. J., Petigura, E. A., Howard, A. W., et al. 2017, "The California-Kepler Survey. III. A Gap in the Radius Distribution of Small Planets", AJ, 154, 109

Fulton, B. J. & Petigura, E. A. 2018, "The California Kepler Survey VII. Precise Planet Radii Leveraging Gaia DR2 Reveal the Stellar Mass Dependence of the Planet Radius Gap," arXiv:1805.01453

Furlan, E. et al. 2017, "The Kepler Follow-up Observation Program. I. A Catalog of Companions to Kepler Stars from High-Resolution Imaging," AJ, 153, 71

Furlan, E. et al. 2018, "The Kepler Follow-up Observation Program. II. Stellar Parameters from Medium- and High-resolution Spectroscopy," ApJ, 861, 149

Gaudi, B. S. et al. 2017, "A giant planet undergoing extreme-ultraviolet irradiation by its hot massive-star host," Nature, 546, 514

Hirsch, L. A., Ciardi, D. R., Howard, A. W., et al. 2017, "Assessing the Effect of Stellar Companions from High-resolution Imaging of Kepler Objects of Interest", AJ, 153, 117

Kraus, A. et al. 2016,"The Impact of Stellar Multiplicity on Planetary Systems. I. The Ruinous Influence of Close Binary Companions," AJ, 152, 8

Morton, T. D. 2012, "An Efficient Automated Validation Procedure for Exoplanet Transit Candidates", ApJ, 761, 6

Raghavan, D., McAlister, H. A., Henry, T. J., et al. 2010, "A Survey of Stellar Families: Multiplicity of Solar-type Stars", ApJS, 190, 1

Stassun, K. G., Corsaro, E, Pepper, J., Gaudi, B. S. 2018, "Empirical Accurate Masses and Radii of Single Stars with TESS and Gaia", AJ, 155, 22

Stefansson, G., Li, Y., Mahadevan, S., et al. 2018, "Diffuser-assisted Photometric Follow-up Observations of the Neptune-sized Planets K2-28b and K2-100b" ArXiv e-prints, arXiv:1807.04420

Sullivan, P. W., Winn, J. N., Berta-Thompson, Z. K., et al. 2015, "The Transiting Exoplanet Survey Satellite: Simulations of Planet Detections and Astrophysical False Positives", ApJ, 809, 77

Ziegler, C. et al. 2018, "Measuring the Recoverability of Close Binaries in Gaia DR2 with the Robo-AO Kepler Survey,"  arXiv: 1806.10142